\documentclass[twocolumn,a4paper,superscriptaddress]{revtex4-1}
\usepackage[utf8]{inputenc}
\usepackage[english]{babel}
\usepackage{amsmath}
\usepackage{amssymb}
\usepackage{amsfonts}
\usepackage[dvips]{color}
\usepackage{graphicx}
\usepackage{hyperref}

\newcommand{\mathd}{\mathrm{d}}
\newcommand{\mathe}{\mathrm{e}}

\begin{document}
\title{\bf Do deviations of neutron scattering widths distribution from
the Porter-Thomas law indicate on failure of the Random Matrix
theory?}
\author{Oleg V. Zhirov}
\email{O.V.Zhirov@inp.nsk.su}
\affiliation{Budker Institute of Nuclear Physics of SB RAS -
Novosibirsk, Russia}
\affiliation{Novosibirsk State University - Novosibirsk, Russia}
\affiliation{Novosibirsk Technical University - Novosibirsk,Russia}

\begin{abstract}
Deviations of neutron scattering width distributions from the Porter-Thomas law 
due to resonances overlapping are calculated in the extended framework of  the 
random matrix approach.
\end{abstract}


\maketitle

\section{Introduction} 
Large irregularities found in spectra in elastic neutron resonant scattering on nuclei 
are widely accepted as a signature of strong quantum chaos in the internal 
dynamics of heavy nuclei. The random matrix theory (RMT) \cite{Mehta04} that  presents an
approach of extreme chaos is  believed to be consistent with the most of nuclear 
data, nicely reproducing both the nearest neighbor levels spacing  distributions and the 
distributions of the resonances widths.

However, recently it was stated \cite{Koehl12}  that a more careful analysis of old 
existing datasets \cite{Koehl11}, as well as the new more precise data 
\cite{Koehl10} on the neutron resonant scattering  indicate on noticeable deviations 
of widths distribution from the Porter-Thomas (PT) law\cite{Porte56} . Since the latter was 
often accepted as a basic prediction of RMT, these deviations  were considered 
in \cite{Koehl12} as a ground to doubt the validity of RMT approach as itself.

These deviations can be treated as an influence of different other \textit{non-chaotic}
mechanisms (e.g. as the influence of peak of neutron strength function 
\cite{Weide10}, parent-daughter correlations\cite{Volya11},  nonstatistical 
$\gamma$-decays \cite{Volya15}) that can cause deviations from ``true chaos'' expectations.
Meanwhile, one should also remind \cite{Celar11}, that even within the frameworks of RMT  the 
\textit{pure} Porter-Thomas law\cite{Porte56} is valid  for \textit{nonoverlapping} resonances only. 
Actually, already the level and widths joint statistics found in the extended RMT approach \cite{Sokol89} indicated on the mutual influence between 
widths and level spacings of resonances, dependent on the  overlapping parameter defined as
\begin{equation}\label{eq:def_kappa}
          \kappa=\left\langle \Gamma \right\rangle /D,
\end{equation}
where $\left\langle \Gamma \right\rangle$ is the mean resonance width and 
$D$ is the mean level spacing.

Despite to the \textit{joint} statistics for a set of  overlapping neutron resonances with 
their energy levels and widths distributions in the frameworks of RMT is known 
for a long time ago \cite{Sokol89} (see also \cite{Fyodo97}, and references therein),
the joint widths and levels distribution obtained in \cite{Sokol89} can hardly be directly applied to 
available experimental data (like in \cite{Koehl12}) in order to test whether 
the deviations from PT law are compatible with expectations within RMT.
Instead of joint distribitions dependent on all the widths and level positions 
one needs  for more practical, so called ``inclusive'' width distribution, e.g.  the distribution
of width for some \textit{given} resonance, that is averaged over widths and positions of all 
the rest ones. However, this distribution was not yet obtained in the clear explicit form, 
that cause some challenge \cite{Shche12,Fyodo15,Bogom16} to get corresponding analytical 
estimates.

Another problem that deserves to be mentioned is also the notorious \textit{``semi-circle'' law},
the inherent prediction of RMT for global distribution of levels, that has no common with actual level
distributions in nuclei. This implies that only a small part of levels far enough from boundaries 
and located in center of the ``semi-circle'' is relevant for comparison with data. A way
to cure this problem imposing in the level space \textit{periodic boundary conditions} was 
proposed by Dyson \cite{Dyson62,Dyson62a,Dyson62b} with the approach based on      
\textit{circular ensembles} of random matrices. Within this approach all the levels are ``central''
with same level density in contrast to traditional RMT. 

One may expect that local predictions
of both approaches, for \textit{central} levels in RMT and levels in circular ensembles,
taken at the same mean level density, should be the same in the limit $N\to\infty$, 
where $N$ is the total number of levels. This equivalence of RMT and circular ensembles 
in application to particular systems (e.g. nuclei) is actually based on the implicit assumption, 
that  the relevant quantities are formed locally, inside the range of levels energy, 
where the level density  can be considered to be approximately constant.

This idea were actually employed in the work \cite{Shche12} where an unphysical \textit{global} 
``semi-circle'' law was simply replaced by  constant levels density spanned 
over infinite range of energy. In turn, the form of \textit{local} mutual influence of resonances
(including its particular pairwise structure) was inherited from RMT \cite{Sokol89}.

An important gain of the work \cite{Shche12} is that \textit{local} resonance-to-resonance 
interaction form coming from extreme chaos assumptions is sepatated from the model-specific 
\textit{global} distribution of levels over energy: accepting the former one may replace the 
latter by some more realistic one.

Intuitively this idea is supported by an analogy between the level-width system and 
log gas of charged particles (see, e.g. \cite{Dyson62,Dyson62a,Dyson62b,Somme88,Haake92a,Forre16}). 
Moreover one can expect for some parameter like Debye screening length that separates 
the particular interactions of nearest neighbors from the mean field influence of far particles.

Unfortunately, the estimates of width distribution presented in \cite{Shche12} were restricted by  
one more simplifying ansatz, that takes the levels be ``pinned'' at fixed positions equidistantly in 
energy. This ansatz simplify calculations substantially, but it appears to be too crude for nearest 
neighbors contribution. In fact, the pairwise resonances interaction is strongly enhanced at 
their small separations, and an account for variations of spacings between the resonance and 
its nearest neighbors requires a more accurate estimate.

In this work we calculate the width distribution  with an explicit integration over
small spacings between the resonance and its two nearest neighbors, while the contribution
of the rest far resonances will be estimated within the equidistant approach used in \cite{Shche12}.
The result is presented in a simple analytical form suitable for comparison with experiments.
In order to verify the estimate we develop a numerical approach that allows to simulate the 
statistics of levels and width, suitable for study of RMT predictions in a clear transparent way.
We demonstrate that our analytical estimate is in a reasonable agreement with numerical simulations.

In Sect.\ref{sec:Scales} we explore the extended RMT expression \cite{Sokol89} for
joined neutron resonances widths and levels statistics, and segregate \textit{local} 
pairwise resonance interactions apart a \textit{global} potential that cause the level 
density distribution to be ``semi-circle''. This allows to introduce a simple model that is free
from the notorious semi-circle law, that can be very useful in numerical simulations and analytical
estimates.  To get a more physical insight in Sect.\ref{sec:Num} we perform the simulations and found
that for the overlapping parameter $\kappa\lesssim0.3$ (far from the width collectivization border
\cite{Sokol92} at $\kappa\sim 1$). Then, with experience of the numerical simulations we get
an analytical estimates that agree  well with the numerical data. 

\section{Different scales in RMT: domain of chaos and the semicircle law}\label{sec:Scales}

In the work \cite{Sokol89} the standard random matrices theory \cite{Mehta04} was 
extended to unstable states.  In the case of one open channel the joint distribution for 
energies  $\vec{E}=\{E_n\}$ and widths $\vec{\Gamma}=\{\Gamma_n\}$ ($n=1,\ldots,N$)
were found in the form:
\begin{eqnarray}
&&\mathd{\cal P}(\vec{E}; \vec{\Gamma})=
   C_N\cdot  \left( \frac{1}{a} \right)^{\frac{N(N+1)}{2}}
   \left(\frac{N}{2\pi\eta}\right)^{\frac{N}{2}}  \times
\nonumber\\
&&\quad \times  \prod_{m<n} 
          \frac{
                   (E_m-E_n)^2+  \frac{1}{4}(\Gamma_m-\Gamma_n)^2}
                   {\sqrt{(E_m-E_n)^2+\frac{1}{4}(\Gamma_m+\Gamma_n)^2}
         }\nonumber\\
&&\quad\times \exp\left[-N
     \left( \frac{1}{a^2}\sum_n E_n^2+
             \frac{1}{2a^2}\sum_{m<n}\Gamma_m\Gamma_n+
             \frac{1}{2\eta}\sum_{n}\Gamma_n
     \right)
\right]\nonumber\\
&&\quad\times\prod_n \mathd E_n \frac{\mathd\Gamma_n}{\sqrt{\Gamma_n}}.
\label{eq:distSV}
\end{eqnarray}
where, according to \cite{Sokol89,Mehta60}, the normalization constant
\begin{equation}\label{eq:CN}
C_N = N^{\frac{N(N+1)}{4}}  
2^{\frac{N(N-1)}{4}}\frac{1}{N!}
\left[\prod_{n=1}^{N}\Gamma(n/2)\right]^{-1}.
\end{equation}
Here the parameter $a$ controls the energy range
occupied by levels, while $\eta$ is related with the mean level width 
$\langle\Gamma\rangle$ by condition  $\eta=N\langle\Gamma\rangle$.

In the limit $N\to\infty$ all the levels are distributed inside the energy
interval $(-a,a)$ by a famous ``semi-circle'' law:
\begin{equation}\label{eq:semi-circle}
 \frac{\mathd n}{\mathd E}=N\frac{2}{\pi a}\sqrt{1-(E/a)^2}.
\end{equation}

The chaotic nature of RMT manifests itself in the level spacing statistics and \textit{local correlations}
of levels, while the global envelope (\ref{eq:semi-circle}) is nothing but kinematical restrictions
of Wigner RMT model, that consider $N$ initially degenerated levels with $E=0$ 
spread over ``semi-circle'' by finite random interactions. %
\footnote{These restrictions can be lifted, e.g. in the Dyson's circular ensembles\cite{Dyson62},
imposing in the level energy space periodic boundary conditions.} 
Obviously, the width distribution of resonances depends on whether the 
resonance is placed at the center of the ``semi-circle'' or near its boundary at $E=\pm a$. %

In order to decrease the influence of bounds at $E=\pm a$, one may consider the resonances lying
inside a domain
\begin{equation}\label{eq:Delta-c}
     -\Delta < E < \Delta
\end{equation}
with  $\Delta\ll a$, where the local mean level spacing $D=\pi a/2N$ is approximately constant:
\begin{equation}
D^{-1}=\frac{\mathd n}{\mathd E}=N\frac{2}{\pi a}\left(1-O(\Delta^2/a^2) \right)
           \approx  \mathrm{const}.
\end{equation}
In addition, one should also require that size of this domain $\Delta$ is still large compared to any 
correlation length. \\

Due to symmetry of the distribution (\ref{eq:distSV}) around  $E=0$, it is convenient to define the index 
range as $m,n\in (-L,\ldots,0,\ldots,L)$, where  $N=2L+1$. Moreover, due to symmetry 
of the Eq.(\ref{eq:distSV})  with respect to any permutation inside the set of pairs 
$\left\lbrace (E_n,\Gamma_n)\right\rbrace$, without any loss of generality,
one can consider the set of pairs $(E_n,\Gamma_n)$  be ordered in energy:
$E_m<E_n$ for any $m<n$, so the pair with $n=0$ appears near the maximum of the 
distribution (\ref{eq:semi-circle}) \footnote{Note, that in the case of integration over ordered
energies $E_m<E_n$ for any $m<n$ the normalization constant (\ref{eq:CN}) is changed by 
$C_N \to N! C_N $ }.\\

Now, let us note that all the local properties including  levels and widths distribution inside  the domain 
\ref{eq:Delta-c} is governed only by the two parameters, by
the mean level spacing $D$ and the mean width $\langle \Gamma \rangle$. 
Using $D$ as a natural energy scale, one may introduce dimensionless variables
\begin{equation}\label{eq:eps.x}
  \varepsilon_n=E_n/D, 
  \qquad x_n=\Gamma_n/\kappa D =\Gamma_n/\langle\Gamma_n\rangle.
\end{equation}
that emphasize the scaling properties of levels system inside the central domain (\ref{eq:Delta-c}).

With these notations the joint distribution (\ref{eq:distSV}) can be rewritten as
\begin{eqnarray}
&& \mathd \bar{\mathcal{P}}(\vec{\varepsilon}; \vec{x})
     =\bar{C}(N)\times\nonumber\\
&&\qquad\times
              \left\lbrace
                  \prod_n \exp\left(-\frac{\pi^2}{4N}\varepsilon_n^2\right)
                              \times\exp\left(-\frac{1}{2}x_n \right)
              \right\rbrace  \label{eq:P1} \\
&&\qquad\times
                 \prod_{m<n}
                    \frac{(\varepsilon_m-\varepsilon_n)^2+
                             \frac{\kappa^2}{4}(x_m-x_n)^2}
                           {\sqrt{(\varepsilon_m-\varepsilon_n)^2+
                                      \frac{\kappa^2}{4}(x_m+x_n)^2}
                           } 
                           \label{eq:P2} \\
&&\qquad\qquad 
                \times
                    \exp\left(-\frac{\pi^2\kappa^2}{8N} x_m x_n
                    \right)
                    \prod_n \mathd \varepsilon_n  \frac{\mathd x_n}{\sqrt{2\pi x_n}}.
               \label{eq:P3}
\end{eqnarray}
Here the normalization constant 
\begin{equation}\label{eq:NormEpsX}
      \bar{C}(N)=(\pi/2)^{N(N+1)/2} 2^{N(N-1)/4}\prod_{n=1}^N \left[\Gamma(n)\right]^{-1}
\end{equation} 
is defined here by the integration over the \textit{ordered} domain 
\[\varepsilon_{-L}<\varepsilon_{-L+1}<\ldots<\varepsilon_0<\ldots
   <\varepsilon_{L-1}<\varepsilon_{L}.\]

\section{The analog statistical mechanics problem.}

It is very instructive to examine the joint distribution of $N$ levels and their widths in terms of 
an analog statistical mechanics problem, as the Boltzmann distribution of $N$ interacting 
particles that are posed into a thermal bath  \cite{Dyson62,Haake92a} (see also the textbook 
\cite{Mehta04}). In this way the joint probability distribution (\ref{eq:P1})-(\ref{eq:P3}) 
can be considered as some partition function 
\begin{equation}\label{eq:PartSum}
  \mathd \bar{\mathcal{P}}(\vec{\varepsilon}; \vec{x})=
    \mathcal{Z}^{-1}\exp(-U (\vec{\varepsilon}; \vec{x})/T)
    \prod_n \mathd \varepsilon_n  \frac{\mathd x_n}{\sqrt{2\pi x_n}}
\end{equation}
In statistical mechanics this is a typical starting point to deduce the properties of the system, 
e.g. single particle distributions, particles correlations, critical phenomena, etc. 
It is also very convenient for computer simulations (see, e.g. many examples in \cite{Binder86}), 
that allows to obtain the distribution directly ``\textit{from the first principles}'', without any further 
approximations in the partition function, that are unavoidable in analytical study. 
In turn, the analytical study, enhanced by the analogy with a condensed matter problem, 
can provide us better understanding the physical properties of
the system. Below we use both of them, in order to justify numerically the
assumptions that underlies the analytical approach used below, as well as to test the final results 
with numerical experiments.\\
 
Let us consider the analog statistical problem in more details.
Without any loss of generality hereafter one can put  ${T=1}$. 
For further convenience one can change the variables 
$x_n \to \xi_n^2$, that turns the Porter-Thomas distribution into a more usable nonsingular form:
\begin{equation}\label{eq:PT-xi}
   \exp(-x/2)\frac{\mathd x}{\sqrt{2\pi x}}\to\exp(-\xi^2/2)\frac{\mathd \xi}{\sqrt{2\pi}},
\end{equation}
where the range of $\xi_n$ is defined as $\xi_n\in (-\infty,\infty)$.
 
Now we have a statistical system of $N$ particles in two-dimensional space $(\varepsilon,\xi)$.
According to (\ref{eq:PartSum}) and (\ref{eq:P1})-(\ref{eq:P3}) the potential energy 
$U(\vec{\varepsilon},\vec{\xi})$ can be splitted into three terms
\begin{equation}\label{eq:Upot}
 U(\vec{\varepsilon},\vec{\xi})=U^{g} (\vec{\varepsilon}) + U^{PT}(\vec{\xi}) + 
 U^{int}(\vec{\varepsilon},\vec{\xi}).
\end{equation}
where the potentials 
\begin{equation}\label{eq:Ug}
   U^{g}(\vec{\varepsilon} )=\sum_{n=-L}^{L}\frac{\pi^2}{4N}\varepsilon_n^2
\end{equation}
and 
\begin{equation}\label{eq:U-PT}
    U^{PT} (\vec{\xi} ) =\sum_{n=-L}^{L}\frac{1}{2}\xi_n^2
\end{equation} 
confine the system along $\varepsilon$ and $\xi$ directions, respectively.

The mutual particle to particle interactions are repulsive and have a pairwize form:
\begin{eqnarray}\label{eq:Uint}
&&U^{int}(\vec{\varepsilon},\vec{\xi})=
     \frac{1}{2}\sum_{m\neq n} U^{int}(\varepsilon_m-\varepsilon_n,\xi_m,\xi_n)=
\nonumber\\
&&\quad =-\frac{1}{2}\sum\limits_{m\neq n}\log  
                     \frac{(\varepsilon_m-\varepsilon_n)^2+
                        \frac{\kappa^2}{4}(\xi_m^2-\xi_n^2)^2}
                               {\sqrt{(\varepsilon_m-\varepsilon_n)^2+
                        \frac{\kappa^2}{4}(\xi_m^2+\xi_n^2)^2}
                     } \label{eq:Uint1}\\
&&\qquad+\frac{1}{2}\sum\limits_{m\neq n}
                                                      \frac{\pi^2\kappa^2}{8N}\xi_m^2\xi_n^2.
\label{eq:Uint2}
\end{eqnarray}
The sum  $(U^g(\varepsilon)+U^{PT}(\xi))$ presents a global potential well.
In the limit $N\gg 1$ this well is stretched along the $\varepsilon$-direction, 
so the system forms a chain of  particles lined along the $\varepsilon$-axis. 

In the limit $\kappa\to 0$ the particle-particle interaction terms (\ref{eq:Uint1})-(\ref{eq:Uint2}) 
turns into a much simpler form
\begin{equation}\label{eq:Umn2dim}
       U^{int}_{m n}=-\log|\varepsilon_m-\varepsilon_n|. 
\end{equation}
that describes\cite{Dyson62a} electrostatic repulsion between particles of unit charge in 
two-dimensional electrodynamics. 
As a balance between the global contracting potential $U^g(\varepsilon)$ and repulsive 
interparticle interaction terms $U^{int}_{mn}$, in presence of the thermal bath with $T=1$ 
the density of particle distribution takes in the limit $N\gg 1$  the ``semi-circle'' form
\begin{equation}\label{eq:s-circ_eps}
 \frac{\mathd n}{\mathd\varepsilon}=\sqrt{1-(2\varepsilon/N\pi)^2}.
\end{equation}
where the semicircle radius $\bar{a}\to \pi N/2$ grows with $N$  linearly.
In turn, the spacing distribution of neighbor particles is known to be close to the Wigner surmise
\begin{equation}\label{eq:WSurmise}
      \frac{\mathd w}{\mathd s} = \frac{1}{2\bar{s}}\exp\left(-\pi s^2/4\bar{s}^2\right).
\end{equation}
where the mean spacing $\bar{s}=(\frac{\mathd n}{\mathd \varepsilon})^{-1}$.\\

The analog statistical system is in fact the Wigner crystal\cite{Wigne34} of charged particles 
that is posed into the global potential $U^g(\varepsilon)=(\pi^2/4N)\varepsilon^2$. The role of 
the potential $U^g(\varepsilon)$ is two-fold:
\begin{itemize}
\item  it provides the boundary conditions that hold the particles together 
preventing the system from expansion due to electrostatic repulsion of  
particles, and

\item  it deforms the crystal, modulating  the particle density by a (rather unphysical) 
semi-circle law form (\ref{eq:s-circ_eps}).
\end{itemize}

Apart from the long-range \textit{global} particle density modulation there are 
\textit{local} particle correlations due to their mutual interactions (\ref{eq:Uint1}),(\ref{eq:Uint2}). 
This correlation effects involve mainly nearest partners, while the influence of \textit{distant} 
partners acts effectively as some mean field, and is generally responsible for the global modification of 
particle distribution. If the corresponding correlation length 
(the ``Debye screening'' parameter) $\lambda_{\varepsilon}\ll \bar{a}$,
all the correlation effects are expected to be \textit{local} phenomena, that are governed by 
the \textit{local} particle density and practically independent of the crystal boundary conditions.
Then the condition (\ref{eq:Delta-c}) turns into
\begin{equation}\label{eq:DeltaEps}
        -\Delta_\varepsilon<\varepsilon<\Delta_\varepsilon, 
\end{equation}
where $\Delta_\varepsilon$ obeys the restrictions
$\Delta_\varepsilon\gg \lambda_{\varepsilon}$ and
$\Delta_\varepsilon\ll \bar{a}$. The latter restriction provides, that in this domain the mean particle
spacing $\bar{s}$ is approximately constant
\begin{equation*} 
(\bar{s})^{-1}=1- \frac{1}{2}(2\varepsilon/N\pi)^2 +\ldots\approx 1,
\end{equation*}
and the \textit{mean positions} of particles inside the domain (\ref{eq:DeltaEps}) 
can be considered as equidistant  \cite{Shche12}.

\subsection{A map on a circle: a way to get rid of semicircle law.}\label{sec:MapOnCirc}

A radical way to get rid of the semicircle law for level energy distribution is to map the energies of particles
onto the unit circle as it was proposed by Dyson \cite{Dyson62,Dyson62a}:
\begin{equation}\label{eq:map2circle}
  \varepsilon\in (-L,\ldots,0,\ldots,L)\to \theta\in (-\pi,\ldots,0,\ldots,\pi).
\end{equation}
In fact, Dyson had considered in\cite{Dyson62,Dyson62a,Dyson62b} \textit{circular ensembles} of random 
\textit{unitary} matrices for \textit{closed} systems of stable levels with \textit{zero} widths. Extention
of circular ensembles to \textit{open systems} was given in later works \cite{Fyodo00,Fyodo01} 
and in the recent
work \cite{Killi16} the join distribution for levels and width was obtained for any $\beta$, that
includes  $\beta=1$ ($T$-invariant), $\beta=2$ ($T$-noninvariant)   and $\beta=2$ (symplectic) ensembles.

Here we restrict ourselves by the $T$-invariant case, and consider the RMT joint distribution
(\ref{eq:distSV}) with minimalistic modifications. Using the map (\ref{eq:map2circle}) 
we we make the substitution
\begin{equation}
(\varepsilon_m-\varepsilon_n)^2 \to \left|\mathe^{i\frac{2\pi}{N}\varepsilon_m}-
                      \mathe^{i\frac{2\pi}{N}\varepsilon_n}\right|^2 \left( \frac{N}{2\pi}\right)^2
\end{equation}
into eq.(\ref{eq:Uint1}). We refer this model as \textit{circularly modified RMT} (cmRMT) .

Geometrically, in this model the particles (levels) are placed on the surface of the unit 
radius cylinder. Their positions in $\{\varepsilon_n\}$ are mapped onto the cylinder directrix, while the positions
in  $\xi_n$ are mapped along cylinder generatrices. Along $\xi$ particles are kept by the potential
(\ref{eq:U-PT}) and particle-particle interaction terms
\begin{widetext}
\begin{equation}
U^{int}(\vec{\varepsilon},\vec{\xi})
 =-\frac{1}{2}\sum\limits_{m\neq n}\log  
                     \frac{\left|\mathe^{i\frac{2\pi\varepsilon_m}{N}}-
                                           \mathe^{i\frac{2\pi\varepsilon_n}{N}}\right|^2 \left( \frac{N}{2\pi}\right)^2+
                        \frac{\kappa^2}{4}(\xi_m^2-\xi_n^2)^2}
                               {\sqrt{\left|\mathe^{i\frac{2\pi\varepsilon_m}{N}}-
                                                     \mathe^{i\frac{2\pi\varepsilon_n}{N}}\right|^2 \left( \frac{N}{2\pi}\right)^2+
                        \frac{\kappa^2}{4}(\xi_m^2+\xi_n^2)^2}
                     }
+\frac{1}{2}\sum\limits_{m\neq n}
                                                      \frac{\pi^2\kappa^2}{8N}\xi_m^2\xi_n^2.
\end{equation}
\end{widetext}
Let us stress, that in cmRMT model all the levels are placed homogeneously on the compact unit circle. 
The ``global'' potential (\ref{eq:Ug}) that had  kept levels inside the range 
$\varepsilon\in \{-\frac{\pi N}{2},-\frac{\pi N}{2} \}$ in the original RMT \cite{Sokol89} is 
here irrelevant and should be omitted. Obviously, in the  limit $N\to\infty$ predictions of of cmRMT and 
RMT models coincide.

Due to periodicity along the axis $\varepsilon$  all the levels have the same mean spacing
and the same width distribution. 
This is very advantages in numerical simulations of the system,  since the statistics can be 
substantially improved by averaging the width distribution over \textit{all the levels}, 
rather than over a small subset of levels (\ref{eq:DeltaEps})  around the center of the semicircle 
(\ref{eq:s-circ_eps}) in the original RMT.

However, one should admit that the original RMT appears to be more convenient to get analytical 
approach for the level width distribution.

\subsection{Non-overlapping resonances ($\kappa=0$).}

In the limit of non-overlapping resonances ($\kappa=0$)  width and level distributions decouples,
with the same Porter-Thomas width distribution  for all the levels. The level energies distribitions
are
\begin{equation}
 \propto \prod_{m, n} |\varepsilon_n -\varepsilon_m| \times
\exp\left\lbrace-\sum_n \frac{\pi^2}{4N}\varepsilon_n^2\right\rbrace
\end{equation}
for RMT model, while cmRMT model the distribution
\begin{equation} 
 \propto  \prod_{m, n} \left|\mathe^{i\frac{2\pi}{N}\varepsilon_m}-
                       \mathe^{i\frac{2\pi}{N}\varepsilon_n}\right|
\end{equation}
coinsides with corresponding distribution in \cite{Dyson62} for Dyson's $T$-invariant circular random 
ensemble.

\subsection{Overlapping resonances ($\kappa\neq 0 $).}

The Breight-Wigner widths of neutron scattering resonances are in practice finite, and the resonances
can overlap: $\kappa=\langle \Gamma \rangle/D >0$. 

In terms of the \textit{analog} statistical model, at $\kappa\neq 0$ 
the particle to particle interactions, given by eqs.(\ref{eq:Uint1}-\ref{eq:Uint2})
depend not only on the separations along the axis $\varepsilon$ 
but on the positions in $\xi$ as well. This extra dependence on $\xi$ causes the deviations
in particle distribution along $\xi$ from the \textit{normal law} that is provided 
by the potential $U^{PT}(\xi)=\xi^2/2$ (\ref{eq:U-PT}):  the normal law becomes valid only in the limit 
$\kappa\to 0$. In terms of the \textit{original} neutron resonant scattering problem 
this fact results in deviations of the neutron scattering width distributions from 
the famous Porter-Thomas law.\\

Let us analyse the interaction terms (\ref{eq:Uint1}-\ref{eq:Uint2}) in more details.

The term (\ref{eq:Uint2}) describes the interactions that causes anticorrelations between
$\xi_m^2$ and $\xi_{n}^2$ for any pair  $(m,n)$. It is curious that these anticorrelations are 
\textit{completely independent} of the pair spacing $|\varepsilon_m-\varepsilon_n|$.
Despite to that each pair contribution \textit{drops} with $N$ as $N^{-1}$, 
for any given $n$-th particle  the net effect coming from all $(N-1)$ its partners 
in the limit $N\to\infty$ remains  \textit{finite}
so the sum over all the partners acts effectively as some mean field: 
$N^{-1}\sum_{m\neq n}\xi^2_m=\langle \xi^2 \rangle+ O(N^{-1/2})$. 
In this way the term (\ref{eq:Uint2}) turns into
\footnote{Note, that an extra factor 2 due to 
double summation cancels the factor $1/2$!
}:
\begin{equation}\label{eq:intMField}
\frac{1}{2}\sum\limits_{m\neq n}^{N}
                                                      \frac{\pi^2\kappa^2}{8N}\xi_m^2\xi_n^2\to
 \frac{\pi^2\kappa^2}{8}\langle \xi^2 \rangle \sum_n \xi_n^2.
\end{equation}
Below we restrict ourselves by the case $\kappa\ll 1$, far from the regime of superradiance
transition\cite{Sokol89} that takes place at $\kappa\sim1$. At small $\kappa$ one has 
$\langle \xi^2 \rangle=1+O(\kappa^2)$, and with an accuracy sufficient for our purposes one can 
safely put hereafter $\langle \xi^2 \rangle=1$.
Finally, this term results in some modification of the potential $U^{PT}(\xi) (\ref{eq:U-PT})$:
\begin{equation}\label{eq:U-PTmod}
     U^{PT}(\xi)\to \tilde{U}^{PT}(\xi)=
     \frac{1}{2}\left(1+\frac{\pi^2\kappa^2}{4}\right)\xi^2
\end{equation}\\

Now, let us consider the mutual particle interactions given by eq.(\ref{eq:Uint1}).
In contrast to (\ref{eq:Uint2}), the contribution of any pair $(m,n)$  in the sum (\ref{eq:Uint1})
 is dependent not only on the particles positions $\xi_m,\xi_n$, but  on their 
$\varepsilon$-spacing $s_{mn}=|\varepsilon_m -\varepsilon_n|$ as well. Expansion in series
over $\kappa\ll 1$ gives:
\begin{eqnarray}
&&\log\left[\frac{s_{mn}^2+ \frac{\kappa^2}{4}(\xi_m^2-\xi_n^2)^2}
                           {\sqrt{s_{mn}^2+
                                      \frac{\kappa^2}{4}(\xi_m^2+\xi_n^2)^2}
                           }\right]
                           =\nonumber\\
&&\qquad\qquad =
    \log(s_{mn})-\frac{3\kappa^2\xi_m^2\xi_n^2}{4s_{mn}^2}+
    \frac{\kappa^2(\xi_m^4+\xi_n^4)}{8s_{mn}^2}
    +\dots.\quad\label{eq:Fmn}
\end{eqnarray}
It is seen, that the second term in the expansion is responsible for anticorrelations between $\xi_m^2$ and 
$\xi_{n}^2$ while the third term causes deviation from the original distribution
due to extra dependence on $\xi_n^4$.  

Both of these contributions are proportional to $s_{mn}^{-2}$
and are enhanced at small spacings $s_{mn}\ll 1$, and their weights become order of unity 
at spacing
\begin{equation}\label{eq:s_c}
        s_{mn} \lesssim \max (\kappa \xi_m^2, \kappa \xi_n^2).
\end{equation}
\begin{figure}[h!]
\begin{center}
\includegraphics[height=3.cm,angle=0]{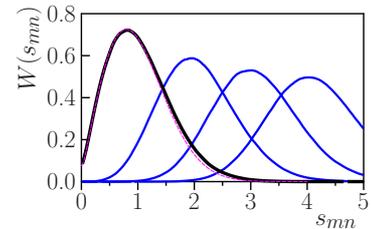}
\end{center}
\caption {\label{fig:gaps_mn} Probability distribution of inter-particle spacing
$s_{mn}=|\varepsilon_k-\varepsilon_l|$ for nearest neighbors $m=n\pm1$ (black line)
and for next to nearest neighbors $m=n\pm(j+1)$, with $j=1,2,3$ (blue lines). 
The dashed magenta line show the modified Wigner surmise (see text for 
explanations). Parameters  used in numerical simulations are $N=321$, $\kappa=0.1$, 
the data are averaged over central levels with $m,n\in (-30,30)$.
}
\end{figure}
In turn, the distribution of spacings $s_{mn}$ is crucially dependent on whether the particles 
$(m,n)$ are nearest neighbors or not. In Fig.\ref{fig:gaps_mn} we illustrate this
by plotting the data taken from numerical simulations for the partition function (\ref{eq:PartSum})
with $U(\vec{\varepsilon},\vec{\xi})$ given by (\ref{eq:Ug})-(\ref{eq:Uint2}) at
$\kappa=0.1$ and $N=321$. The spacings were measured for particles lying
in the center of the crystal with $m,n\in (-30,30)$, that correspond to 
$\Delta_\varepsilon\approx 30\ll \bar{a}\approx 504$. 

One can see that  the distribution of spacings $s_{mn}$ for nearest neighbors 
($m=n\pm 1)$ (black line) at $s_{mn}\ll1$ drops only linearly that being combined with 
expansion (\ref{eq:Fmn}) gives singularity $s_{mn}^{-1}$ resulting in logarithm 
enhancement for contribution of small spacing. 

In contrast, for pairs sandwiched
by one and more particles ($j=1,2,3,\ldots$) the distribution of spacings shown by 
blue lines are highly suppressed at $s_{mn}\to 0$. Note, that in the latter case the 
distributions  are peaked approximately at $s_{mn}\approx (j+1)$ and 
corresponding contribution can be estimated with some reasonable accuracy 
under assumption that the far partners are placed at fixed equidistant positions 
with $s_{mn}=2,3,\ldots$, like it was proposed in\cite{Shche12}. 

While  far partners can be assumed to be placed equidistant, for nearest neighbors one need know
actual distribution of spacings $s_{n,n\pm1}$. Fortunately, it is remarkably close to the Wigner surmise 
(\ref{eq:WSurmise})  slightly modified as
\begin{equation}\label{eq:modWigSurm}
   \frac{\mathrm{d}w}{\mathrm{d}s}=\mathrm{const}
        \sqrt{s^2+(\kappa/2)^2} \exp(-\pi s^2/4).
\end{equation}
(see dashed magenta curve at Fig.\ref{fig:gaps_mn}). Below we use this fact in our analytical estimates.

\section{Width distribution of overlapping resonances}

The resonance width distribution $\mathbb{P}(x,\kappa)$ that we are going to estimate analytically
is related  in the analog statistical model to the distribution $P(\xi)$ of the particle along the axis $\xi$: 
\begin{eqnarray}
    \mathbb{P}(x,\kappa)\mathd x &=&2 P(\xi,\kappa) \mathd \xi{\Big |}_{\xi=\sqrt{x}}\nonumber\\
      &=&P(\xi=\sqrt{x},\kappa)\frac{\mathd x}{\sqrt{x}}
\end{eqnarray}
 In order to minimize the influence of kinematic boundaries in RMT, we consider a central particle
(with $n=0$) equally far distant from  ends of the system:
\begin{eqnarray}\label{eq:WD0}
&&  P(\xi_0,\kappa,L)=
 Z^{-1}\int \mathd s_{-}\,\mathd s_{+}\prod_{n=-L}^{-1}\mathd \xi_n 
\prod_{n'=1}^{L}\mathd \xi_{n'} \times\nonumber\\
&&\quad\times \exp\left[- \sum_{n=-L}^{L} \tilde{U}^{PT}(\xi_n)-
     \sum_{m\neq n} \tilde{U}_2^{int}(\varepsilon_m-\varepsilon_n; \xi_m,\xi_n)
 \right], \nonumber\\
\end{eqnarray} 
where for the central partice we can put  $\varepsilon_0=0$, and for the nearest neighbor 
separations $s_{\pm 1}=\varepsilon_{\pm 1}$, while far partners are taken like in \cite{Shche12}
at equidistant positions:
$\varepsilon_n =n$, for $n=\pm 2,\pm 3,\ldots,\pm L$.

Then the integration over the far partner variables  $\xi_n$, $n=\pm 2,\pm 3,\ldots,\pm L$ can be perfomed by the saddle point method and we get in the limit $L\to\infty$ (see Appendix\ref{calcWD}):
\begin{eqnarray}\label{eq:WD1}
&&P(\xi_0=\sqrt{x},\kappa)=C(\kappa)\times \mathcal{A}(x,\kappa)\times\nonumber\\
&&\quad\times\exp\left\lbrace -\left(1+\frac{\pi^2\kappa^2}{4}\right)\frac{x}{2}\right\rbrace
\times\mathcal{I}_0(x,\kappa)\mathcal{I}_1(x,\kappa)
\end{eqnarray} 
where the contribution coming from far partners is given by the factor
\begin{eqnarray}
\mathcal{A}(x,\kappa)&=&\left(\frac{\sinh(\pi\kappa/2)}{\pi\kappa/2}\right)^2
\frac{\pi\kappa\sqrt{2x(3+x)}/4}{\sinh(\pi\kappa\sqrt{2x(3+x)}/4)}\times\nonumber\\
&&\times\frac{\left[1+3x\kappa^2/8\left(1+x^2\kappa^2/4\right)\right]}{\left(1+x^2\kappa^2/4\right)}
\end{eqnarray}
while the integrals (with $l=1,2$)
\begin{eqnarray}
\mathcal{I}_l = \int\mathd\varepsilon\;\mathd\xi
\frac{\varepsilon^2+\frac{\kappa^2}{4}\left(\xi^2-x\right)^2}
{\sqrt{\varepsilon^2+\frac{\kappa^2}{4}\left(\xi^2+x\right)^2}}\times\nonumber\\
 \mathe^{-\xi^2/2}
\varepsilon^l\exp\left\lbrace-\frac{(1+\pi/4)^2}{\sqrt{\pi}\varepsilon^2} \right\rbrace
\end{eqnarray}
This integrals can be easily be either evaluated  numerically or estimated analyticaly.

The normalization factor $C(\kappa)$ is fixed by a condition 
\begin{equation}\label{eq:Pnorm}
 \int\limits_{-\infty}^{\infty} P(\xi_0)\mathd \xi_0= \int\limits_{0}^{\infty}\mathbb{P}(x)\mathd x =1.
\end{equation}

The effect of resonances interactions on the width distribution is better seen
in the ratio  of $\mathbb{P}(x,\kappa)=P(\xi_0=\sqrt{x},\kappa)/\sqrt{x}$ to 
the Porter-Thomas distribution 
$W_{PT}(x)=\exp(-x/2)/\sqrt{2\pi x}$:
\begin{eqnarray}
&&\Psi(x|\kappa)=\mathbb{P}(x,\kappa)/W_{PT}(x)=\nonumber\\
&&\quad=\pi C(\kappa) \mathcal{A}(x,\kappa)
\exp\left\lbrace -\frac{\pi^2\kappa^2}{8}x\right\rbrace
\mathcal{I}_0(x,\kappa)\mathcal{I}_1(x,\kappa)\nonumber\\
\label{eq:Psi}
\end{eqnarray}
Indeed,  the normalization factor $C(\kappa)$ can also be fixed by the condition
\begin{equation}\label{eq:Normalizatio}
    \int_0^\infty \frac{\mathrm{d}x}{\sqrt{2\pi x}}\; \Psi(x|\kappa)\mathrm{e}^{-x/2}=1.
\end{equation}

\begin{figure}[tbph]
\centering
\includegraphics[width=0.7\linewidth]{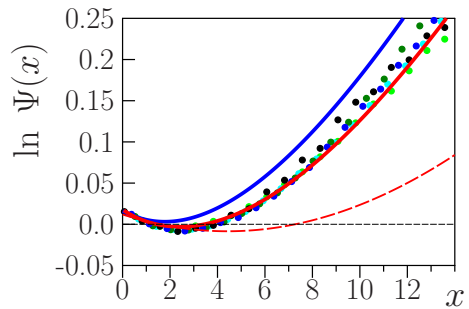}
\caption{$\log\Psi(x|\kappa)$ for $\kappa=0.05$. Red line presents our analytical estimate given
by   eq.(\ref{eq:Psi}). Results of our numerical simulations of the RMT joined probability
distribution  (\ref{eq:distSV}) (see App.\ref{sec:Num}) are shown 
by colored points: $N=3$ (lime), $N=5$ (cyan), $N=9$ (green), $N=11$ (blue) and $N=101$ (black ).
The red dashed lines show results of the equidistant estimate (see text).
The blue line plots recent results obtained by Fyodorov and Savin \cite{Fyodo15}. 
} 
\label{fig:k=0.05}
\end{figure}

Let us start with the case of weak overlap $\kappa=0.05$, see Fig..\ref{fig:k=0.05}. 
 the corrected $\Psi(x|\kappa)$ is compared to results of our numerical 
simulations for direct the RMT joint distribution (\ref{eq:distSV}) (see for details
 Appendix \ref{sec:Num}). In these simulations we perform runs with the number of 
levels varied from $N=3$  to $N=101$. We see that our analytical estimate 
(\ref{eq:Psi}) agrees well with numerical data. The red dashed lines show the expectations 
of the ``equidistant'' ansatz, that assumes that all the resonances are placed on 
equidistant positions, like it was done in  \cite{Shche12}. It is clearly seen that the 
nearest neighbor position fluctuations play a very important role (especially at smaller 
overlapping parameter $\kappa$). We plot also recent results obtained
in the work \cite{Fyodo15}. Despite to the qualitative behavior  is similar, quantitatively 
the effects of the resonance overlapping is a bit overestimated.

\begin{figure}[t!h!]
\begin{center}
\includegraphics[height=3.cm,angle=0]{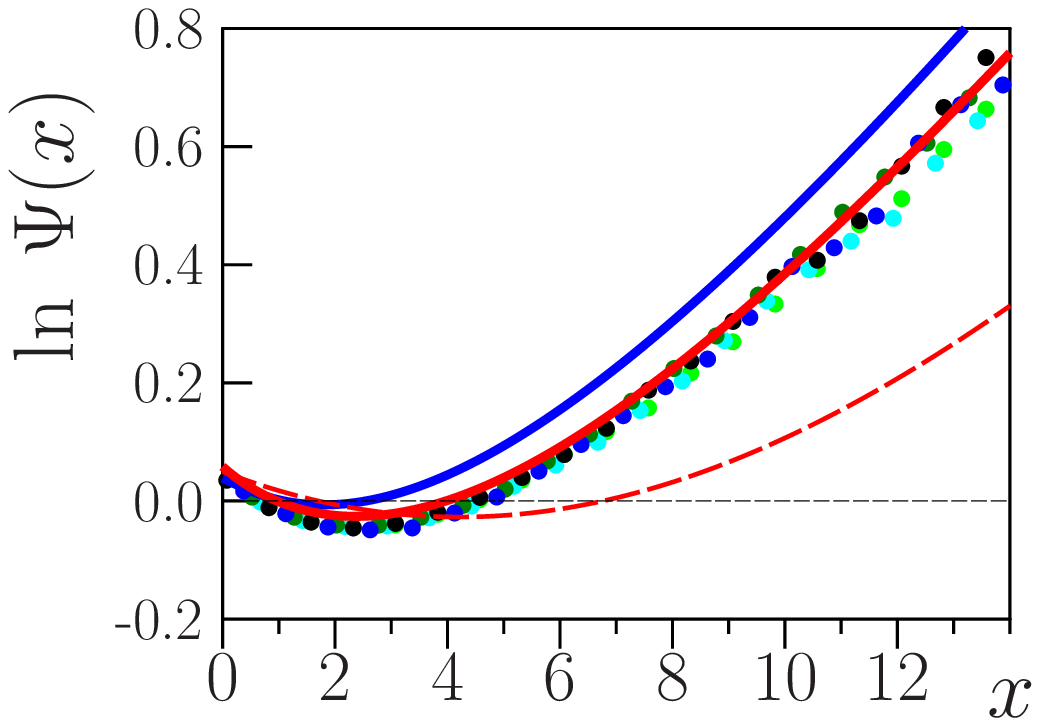}
\includegraphics[height=3.cm,angle=0]{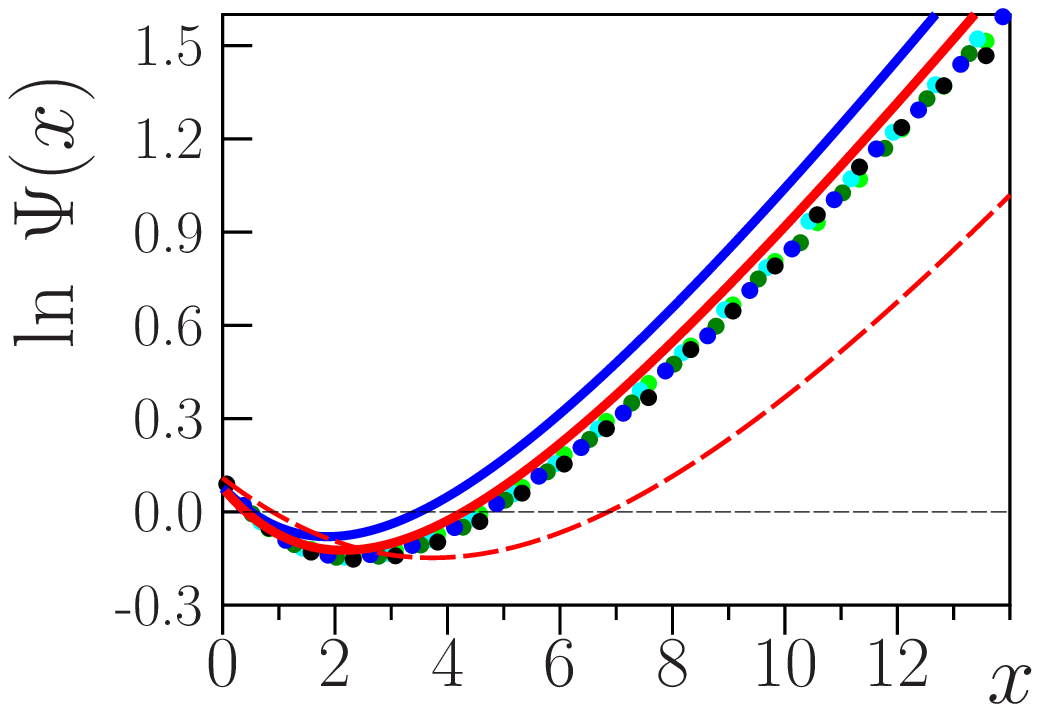}
\end{center}
\caption {\label{fig:Psi}$\Psi(x|\kappa)$ for $\kappa=0.1$ (left) and
$\kappa=0.2$ (right).
Red solid lines present analytical estimate by eq.(\ref{eq:Psi}). 
Colored points show results of obtained from 
numerical simulations of  the joined probability distribution (\ref{eq:distSV})
with $N=321$, $k\in (-30,30)$ (black), $N=161$, $k\in (-10,10)$ (blue),  
$N=81$, $k\in (-8,8)$ (green),  $N=41$, $k\in (-4,4)$ (cyan) and  $N=21$, 
$k\in (-4,4)$ (lime). 
The red dashed lines show results of the equidistant estimate (see text).
Blue lines show the estimates obtained in the recent work \cite{Fyodo15}.
}
\end{figure}

The results for $\kappa=0.1$ and $0.4$ are presented in Fig.\ref{fig:Psi}. 
The estimate (\ref{eq:Psi}) fit the data well up to $\kappa\lesssim0.4$. Note also,
that at higher $\kappa$ the difference between (\ref{eq:Psi}) and analytical estimate
given in \cite{Fyodo15} decreases \footnote{
One can guess that the level-to-level \textit{anticorrelation} effects,
given by  second term in (\ref{eq:Fmn}) are not properly accounted in \cite{Fyodo15}, 
since at the final stage of width distribution estimates (as it was stated by authors of the work) 
the width of all levels except for a given one were taken
to be zero.}.

In the simulations presented here all the data are collected from the levels lying 
either at the center or in the central domain with $k\in (-0.1 N,0.1N)$. 
One can see that already at $N\gtrsim 50$ the width distributions of levels lying in 
the selected domain of $k$ are practically independent of $N$ and reaches its 
asymptotic behaviour expected in the limit $N\to\infty$, provided that $\kappa\lesssim0.5$.

 However, at $\kappa\gtrsim 0.4$ the width distributions of
levels lying even near the center  are found to be dependent on total number 
of levels $N$,   and the asymptotic behavior does not occur  even at 
$N\gtrsim 200-300$. To our mind this indicates on long-range correlations that 
come into play at large overlapping parameter $\kappa$ due to
proximity of super-radiance transition that is expected at  $\kappa=1$ \cite{Sokol89}.

\begin{figure}[t!h!]
\begin{center}
\includegraphics[height=3.cm,angle=0]{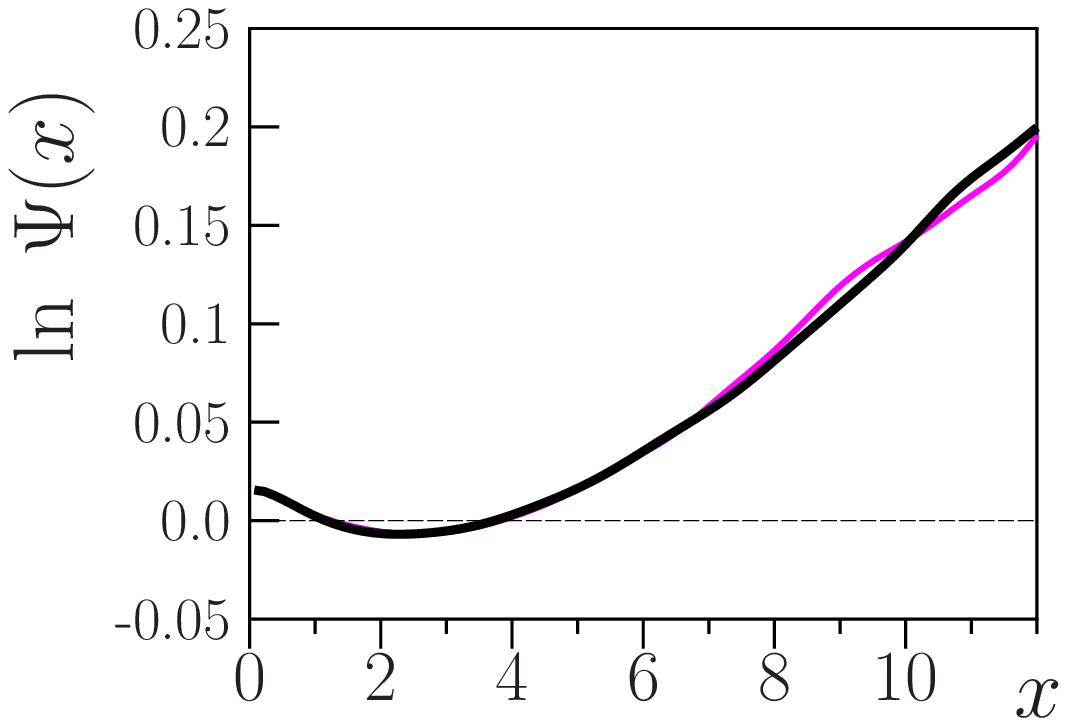}
\includegraphics[height=3.cm,angle=0]{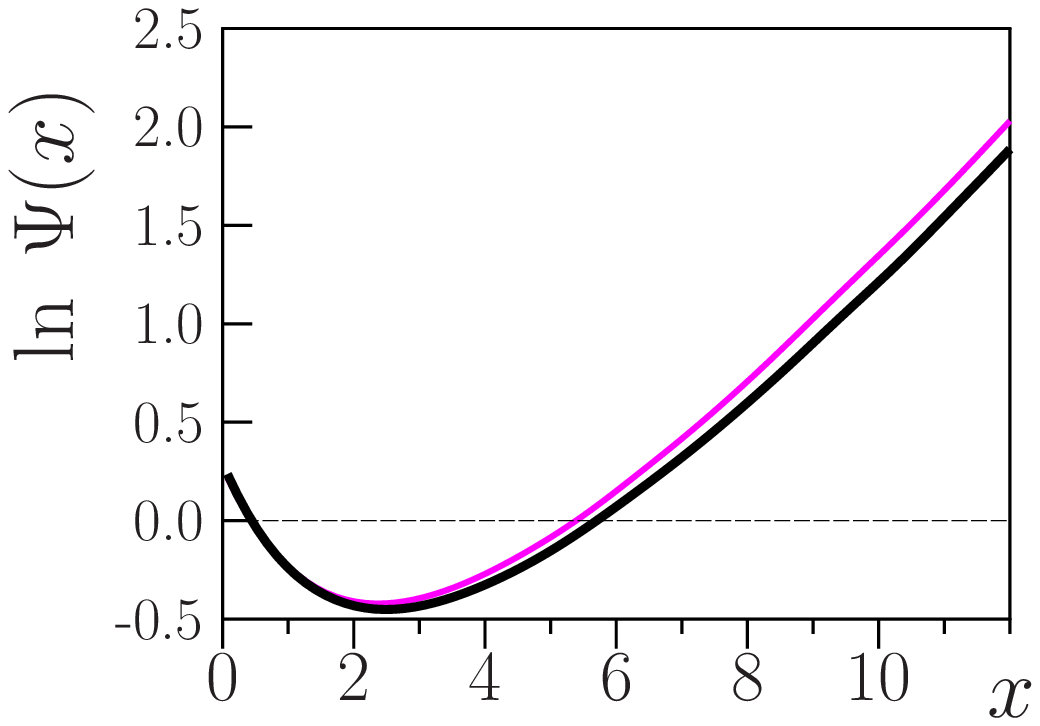}
\end{center}
\caption {\label{fig:cmRMT}$\Psi(x|\kappa)$ for $\kappa=0.05$ (left) and
$\kappa=0.4$ (right).
Black and magenta lines present results of simulation for cmRMT ($N=100$) and  
original RMT ($N=101$). In the latter case the data are taken for levels lying in the
central band $k\in(-11,11)$.
}
\end{figure}
To complete this section let us also consider the predictions of circularly modified RMT (cmRMT),
introduced in the Section \ref{sec:MapOnCirc}. In Fig.\ref{fig:cmRMT} we demonstrate that 
the results are practically identical. However, while with cmRMT  we average data over all the 
levels, in the case of original RMT we must restrict ourselves by a small subset of levels 
lying at the center of the semicircle. 

\section{Conclusions.}
In this paper the deviations of width distributions from the Porter-Thomas law
expected in the extended framework of RMT are explicitly calculated in a clear transparent 
way. Our analytical estimate is in nice agreement with corresponding numerical simulations.
The contribution to deviations comes from two sources: the long range correlations
calculated in equidistant approximation\cite{Shche12} and short-range correlations
found in the nearest neighbors approximation. The role of short-range correlations 
are found especially important at small $\kappa$ where it is logarithmically enhanced.

These deviations from the Porter-Thomas law occur even the dynamics inside the system
is perfectly chaotic, irrespective to possible existence of  contributions coming from
other mechanisms \cite{Weide10,Celar11,Volya11,Volya15} 
that can also cause the deviations from the Porter-Thomas law. Despite to a 
comparative analysis of all this sources is very interesting, in this paper we restrict ourselves 
by effects survive in the limit of the perfect chaos.

\section{Acknowledgments}
I am are very grateful V.V. Sokolov for many years collaboration and many stimulative discussions
that cause my interest to this topic.  This work is
supported by the Ministry of Education and Science of the Russian Federation.


\bibliographystyle{apsrev4-1}
\bibliography{Main}{}
\newpage
\appendix
\begin{widetext}
\section{Calculation of the width distribution (\ref{eq:WD1})}
\label{calcWD}
The distribution of the central particle with $n=0$ along the axis $\xi$ is given 
by the expression 
\begin{eqnarray}\label{eq:WD0a}
&&  P(\xi_0,\kappa,L)=
 {\rm const} \int \prod_{n=-L}^{-1}\mathd \varepsilon_n \mathd \xi_n 
\prod_{n'=1}^{L}\mathd \varepsilon_{n'}\mathd \xi_{n'} 
\quad\times \exp\left[- \sum_{n=-L}^{L} \tilde{U}^{PT}(\xi_n)-
     \sum_{m\neq n} \tilde{U}_2^{int}(\varepsilon_m-\varepsilon_n; \xi_m,\xi_n)
 \right], \nonumber\\
\end{eqnarray} 
Interactions between any two particles are described by a pairwise potential 
(or \textit{link}) :
\begin{equation}\label{eq:U2int}
U_2^{int}(\varepsilon_n-\varepsilon_m;\xi_n,\xi_m)=
-\ln\frac{(\varepsilon_n-\varepsilon_m)^2+\frac{\kappa^2}{4}(\xi_n^2-\xi_m^2)^2}
{\sqrt{(\varepsilon_n-\varepsilon_m)^2+\frac{\kappa^2}{4}(\xi_n^2+\xi_m^2)^2}}
\end{equation}
All the particles in (\ref{eq:WD0a})  with $n\in \left[-L,\ldots,0,\ldots,L\right]$  
are considered as a  subset taken at the center of the global semicircle distribution 
with the radius $a=\pi \mathbb{N}/2$ of much larger system, with total particle number
 $\mathbb{N}\ggg N=2L+1$,
so the particle density inside the chosen subset $\left[-L,\ldots,L\right]$ 
can be considered to be \textit{constant}. 

Let us emphasize in (\ref{eq:WD0a})  the contribution of two nearest neighbors with $n=\pm1$:
\begin{eqnarray}
  P(\xi_0,\kappa,L)&=&{\rm const} \int \mathd \varepsilon_{-1}\mathd \varepsilon_{1}
\mathd \xi_{-1}\mathd \xi_{1} 
\exp\left\lbrace{-\left(1+\frac{\pi^2\kappa^2}{4}\right)
\times\frac{\xi_{-1}^2+\xi_{1}^2+x}{2}}\right\rbrace\times\nonumber\\
&&\quad\times\exp\left\lbrace -U_2^{int}(\varepsilon_{-1};\xi_{-1},\sqrt{x})-
 U_2^{int}(\varepsilon_{1};\xi_{1},\sqrt{x})\right\rbrace 
 \mathcal{F}(\varepsilon_{-1},\xi_{-1};\varepsilon_{1},\xi_{1}|x)
\end{eqnarray} 
Here for the central particle we put $\varepsilon_0=0$, $\xi_0=0$ and 
write out  the integrations over the positions $\varepsilon_{\pm1}$ and
$\xi_{\pm1}$ for two nearest neighbor particles explicitly, while the contributions from the rest 
of far partners
are absorbed into the weight function $\mathcal{F}(\varepsilon_{-1},\xi_{-1};\varepsilon_{1},\xi_{1}|x)$.

The function   $\mathcal{F}(\varepsilon_{-1},\xi_{-1};\varepsilon_{1},\xi_{1}|x)$
can be presented in the form
\begin{equation}\label{eq:calF}
\mathcal{F}(\varepsilon_{-1},\xi_{-1};\varepsilon_{1},\xi_{1}|x)=
\exp\left\lbrace -U_{eff}(\varepsilon_{-1},\xi_{-1};\varepsilon_{1},\xi_{1}|x) \right\rbrace
\end{equation}
where $U_{eff}(\varepsilon_{-1},\xi_{-1};\varepsilon_{1},\xi_{1}|x)$ is the effective mean field,
that acts on the central particle as well as on its two nearest neighbors.

Let us restrict ourselves by equidistant approach for all particles but the central one and its
two neighbors: $\varepsilon_n=n$ for $N\in\left[-L,\ldots,-2,2,\ldots,L\right]$. Then, the
corresponding expression for  $\mathcal{F}(\varepsilon_{-1},\xi_{-1};\varepsilon_{1},\xi_{1}|x)$
takes a form
\begin{eqnarray}
\mathcal{F}(\varepsilon_{-1},\xi_{-1};\varepsilon_{1},\xi_{1}|x)&=&
\int\prod_{n=-L}^{-2}\mathd\xi_n\prod_{n'=2}^{L}\mathd\xi_n'\times
\exp\left\lbrace - \frac{1}{2}\left(1+\frac{\pi^2\kappa^2}{4}\right)
\left[\sum_{n=-L}^{-2}\xi_n^2+\sum_{n'=2}^{L}\xi_{n'}^2\right]\right.-\nonumber\\
&&-\sum_{m=-L}^{-2}U_2^{int}(m;\xi_m,\sqrt{x})-\sum_{m'=2}^{L}U_2^{int}(m';\xi_{m'},\sqrt{x})-
\label{L6}\\
&&-U_2^{int}(\varepsilon_{1}-\varepsilon_{-1};\xi_1,\xi_{-1})-\label{L7}\\
&&-\sum_{m=-L}^{-2}U_2^{int}(\varepsilon_1-m;\xi_m,\xi_1)-
      \sum_{m'=2}^{L}U_2^{int}(\varepsilon_1-m';\xi_m',\xi_1)-\label{L8}\\
&&-\sum_{m=-L}^{-2}U_2^{int}(\varepsilon_{-1}-m;\xi_m,\xi_{-1})-
      \sum_{m'=2}^{L}U_2^{int}(\varepsilon_{-1}-m';\xi_m',\xi_{-1})-\label{L9}\\
&&-\sum_{m=-L}^{-2}\sum_{m'=2}^{L}U_2^{int}(m-m';\xi_m',\xi_m)-\label{L10}\\
&& \left. -\frac{1}{2}\sum_{\underset{m\neq n}{m,n=-L}}^{-2}U_2^{int}(m-n;\xi_m,\xi_n)
-\frac{1}{2}\sum_{\underset{m'\neq n'}{m',n'=2}}^{L}U_2^{int}(m'-n';\xi_m',\xi_n')
       \right\rbrace\label{L11}
\end{eqnarray}
Let us estimate $\mathcal{F}(\varepsilon_{-1},\xi_{-1};\varepsilon_{1},\xi_{1}|x)$ performing integrations
by the saddle point method. For this purpose let us expand links (\ref{L6})-(\ref{L11}) by $\xi$ around
$\xi_n$, $(n=-L,\ldots,-2,2,\ldots,L)$ up to second order terms and explore their contributions into 
the weight function  $\mathcal{F}(\varepsilon_{-1},\xi_{-1};\varepsilon_{1},\xi_{1}|x)$.
\begin{enumerate}
\item Let us start with the important remark, that the great majority of links contributing in 
(\ref{L10}),(\ref{L11}), which number is order of $O(L^2)$ are independent of $x$. Since their expansions
\begin{equation} 
U_2^{int}(m-n;\xi_m,\xi_n)=\frac{1}{2}\ln|m-n|+\kappa^2\cdot O(\xi_m^4,\xi_n^4,\xi_m^2\xi_n^2),
\end{equation}
have no terms of second order in $\xi$, within the accuracy of the saddle point method they gives only some constant factor into the total normalization constant in (\ref{eq:WD0a}.

\item The next important point is  the dependence of
 $\mathcal{F}(\varepsilon_{-1},\xi_{-1};\varepsilon_{1},\xi_{1}|x)$ on $x$. Within the 
 currently used approach  it  comes from the direct pairwise interaction of the central particle ($n=0$) 
 to far partners placed  equidistantly along $\varepsilon$-axis by links (\ref{L6}), which 
 total number is order of $O(L)$. Expansion links in (\ref{L6}) over $\kappa$
 has a form
 \begin{equation} 
    U_2^{int}(n;\xi_n,x) = -\ln\sqrt{n^2+\kappa^2 x^2/4}+ 
    \frac{3}{16} \frac{\kappa^2 x}{n^2(1-\kappa^2 x^2/4n^2)} \xi_n^2+\kappa^2\cdot O(\xi_n^4)
 \end{equation}
 Here we omit terms order of $O(k^2)$  except for those that are multiplied by $x$, since they are enhanced at $x\gg 1$, while main contribution from integration over $\xi_n$ originates from  $|\xi_n|\sim 1$.

\item The links in (\ref{L8}),(\ref{L9}) control the dependence of 
$\mathcal{F}(\varepsilon_{-1},\xi_{-1};\varepsilon_{1},\xi_{1}|x)$ on $\varepsilon_{\pm 1}$, $\xi_{\pm 1}$
and form the distribution of the nearest neighbors. The corresponding corrections to the distribution over
$\xi$ are order of $O(\kappa^2)$ and with accuracy sufficient for further estimates one may restrict himself
by zero order approximation:
\begin{eqnarray} 
U_2^{int}(n-\varepsilon_{\pm 1};\xi_n,\xi_{\pm 1}) &\approx&
-\ln\sqrt{(n-\varepsilon_{\pm 1})^2+\frac{\kappa^2}{4}\xi_{-1}\pm 1}+
\frac{3}{16} \frac{\kappa^2\xi_{\pm 1}^2}
{(n-\varepsilon_{\pm 1})^2\left(1+\frac{\kappa^2\xi_{\pm 1}^4}{(n-\varepsilon_{\pm 1})^2} \right)}
\nonumber\\
&&=-\ln|n-\varepsilon_{\pm 1}|+O(\kappa^2).
\end{eqnarray}
Up to the link (\ref{L7}) the distributions  over $(\varepsilon_{-1},\xi_{-1})$  
and $(\varepsilon_{1},\xi_{1})$  factorize into independent multipliers for each of nearest neighbors. Expanding (\ref{L7}) by powers of $\kappa$ in the limit $\kappa\ll 1$ we get
\begin{eqnarray}
U_2^{int}(\varepsilon_{1}-\varepsilon_{-1};\xi_1,\xi_{-1})&\approx&
-\ln|\varepsilon_{1}-\varepsilon_{-1}|+
\frac{\kappa^2\cdot(\xi_1^4+\xi_{-1}^4-6\xi_1^2\xi_{-1}^2)}{8(\varepsilon_1-\varepsilon_{-1})^2}+
O(\kappa^4)=\nonumber\\
&=&-\ln|\varepsilon_{1}-\varepsilon_{-1}|+O(\kappa^2).
\end{eqnarray}
\end{enumerate}

In summary, within current approach the weight function turns into the product of factors
\begin{eqnarray}
\mathcal{F}(\varepsilon_{-1},\xi_{-1};\varepsilon_{1},\xi_{1}|x)&=&
\prod_{n=-L}^{-2}\sqrt{n^2+\frac{\kappa^2 x^2}{4}}
\prod_{n'=2}^{L}\sqrt{n'^2+\frac{\kappa^2 x^2}{4}}\; 
\exp\left\lbrace -\left(1+\frac{\pi^2\kappa^2}{4} \right)\frac{x}{2}\right\rbrace\times\label{F16}\\
&&\times\int\prod_{n=-L}^{-2}\mathd \xi_n
\exp\left\lbrace -\left(1+\frac{3}{8} \frac{\kappa^2 x}{n^2\left(1+\frac{\kappa^2 x^2}{4 n^2} \right)}
\right)\frac{\xi_n^2}{2}
\right\rbrace\times\label{F17}\\
&&\times\int\prod_{n'=2}^{L}\mathd \xi_{n'}
\exp\left\lbrace -\left(1+\frac{3}{8} \frac{\kappa^2 x}{{n'}^2\left(1+\frac{\kappa^2 x^2}{4 {n'}^2} \right)}\right)\frac{\xi_{n'}^2}{2}
\right\rbrace\times\label{F18}\\
&&\times|\varepsilon_{1}-\varepsilon_{-1}|\times\label{F19}\\
&&\times\prod_{n=-L}^{-2}|n-\varepsilon_{-1}|\prod_{n'=2}^{L}|n'-\varepsilon_{-1}|\times\label{F20}\\
&&\times\prod_{n=-L}^{-2}|n-\varepsilon_{1}|\prod_{n'=2}^{L}|n'-\varepsilon_{1}|=\label{F21}\\
&=&A(x,\kappa,L)B(x,\kappa,L)
\exp\left\lbrace -\left(1+\frac{\pi^2\kappa^2}{4} \right)\frac{x}{2}\right\rbrace\times\nonumber
\label{F22}\\
&&\times|\varepsilon_{1}-\varepsilon_{-1}|g(\varepsilon_{-1},L)g(\varepsilon_{1},L)
\exp\left\lbrace -\frac{\xi_{1}^2+\xi_{-1}^2}{2}\right\rbrace\label{F23}
\end{eqnarray}
Here in the limit  $L\to\infty$ the pre-exponent factors (\ref{F16}) combine into the function
\begin{eqnarray}
A(x,\kappa,L)&=&\left[\prod_{n=2}^{L}\sqrt{n^2+\frac{\kappa^2 x^2}{4}} \right]^2=
\prod_{n=2}^{L}\left[n^2+\frac{\kappa^2 x^2}{4} \right]=\nonumber\\
&=&C(L)\prod_{n=2}^{L}\left[1+\frac{\kappa^2 x^2}{4n^2} \right]
\underset{L\to \infty}{\longrightarrow}
{\rm const}\times \frac{\sinh(\pi\kappa x/2)}{(\pi\kappa x/2)(1+\kappa^2 x^2/4)}
\end{eqnarray}
and integrations in (\ref{F17}),(\ref{F18}) in the limit $L\to\infty$ gives
\begin{eqnarray}
B(x,\kappa,L)&=&\left[\int\prod_{n=2}^{L}\mathd \xi_n
\exp\left\lbrace -\left(1+\frac{3}{8} \frac{\kappa^2 x}{n^2\left(1+\frac{\kappa^2 x^2}{4 n^2} \right)}
\right)\frac{\xi_n^2}{2}\right\rbrace   \right]^2\nonumber\\
&=&\prod_{n=2}^{L}\left[ \sqrt{2\pi/\left(1+\frac{3}{8} \frac{\kappa^2 x}{n^2\left(1+\frac{\kappa^2 x^2}{4 n^2} \right)}
\right)}\; \right]^2=\tilde{C(L)}\prod_{n=2}^{L}\left[ 1+\frac{3}{8} \frac{\kappa^2 x}{n^2\left(1+\frac{\kappa^2 x^2}{4 n^2} \right)}
 \right]^{-1}\nonumber\\
&&\underset{L\to \infty}{\longrightarrow} {\rm const}\times
\left(1+\frac{3k^2 x}{8\left(1+\kappa^2 x^2/4\right)}\right) \frac{\sinh(\pi\kappa x/2)}{\pi\kappa x/2}
\frac{\pi\kappa\sqrt{2x(3+2x)}/4}{\sinh(\pi\kappa\sqrt{2x(3+2x)}/4)}
\end{eqnarray}
Note, that in the limit $\kappa\to 0$ we have $A(x,\kappa,\infty)=B(x,\kappa,\infty)=1$.

Now, the factors (\ref{F19}),(\ref{F20}), (\ref{F21}), where in the equidstant approach 
\begin{equation}\label{key}
g(\varepsilon,L)=\prod_{n=-L}^{-2}|n-\varepsilon| \prod_{n'=2}^{L}|n'-\varepsilon| 
\end{equation}
is an even function with respect to transformation $\varepsilon \to -\varepsilon$. In principle, this function 
should define the nearest neighbor distribution with  due to its interaction to particles with numbers
$n=[-L, \ldots,-2,2,\ldots,L]$. However, the equidistant ansatz is too crude to get a reasonable estimate
for a long range of the distribution, and instead we use below another approach.

It is well known, that the nearest neighbor levels spacing distribution is remarkably close to the Wignern-Dyson surmize 
\begin{equation}\label{Wsurm}
\frac{\mathd w^{WD}}{\mathd s}=\frac{1}{2} \; \exp(-\pi s^2/4)
\end{equation}

According to (\ref{F23}) the joint distribution for two neighbors looks as
\begin{eqnarray}
W(\varepsilon_{-1}, \varepsilon_1; \xi_{-1},\xi_1|x,\kappa)&=& 
{\rm const}\cdot(|\varepsilon_{-1}|+|\varepsilon_{1}|)\times\nonumber\\ 
&&\times g(\varepsilon_{-1},\kappa)\frac{\varepsilon_{-1}^2+\frac{\kappa^2}{4}(\xi_{-1}^2-x)^2}
  {\sqrt{\varepsilon_{-1}^2+\frac{\kappa^2}{4}(\xi_{-1}^2+x)^2}}\exp\left\lbrace -\frac{\xi_{-1}^2}{2}
  \right\rbrace  \times\\
  &&\times g(\varepsilon_{1},\kappa)\frac{\varepsilon_{1}^2+\frac{\kappa^2}{4}(\xi_{1}^2-x)^2}
    {\sqrt{\varepsilon_{1}^2+\frac{\kappa^2}{4}(\xi_{1}^2+x)^2}}\exp\left\lbrace -\frac{\xi_{1}^2}{2}
    \right\rbrace  \\
\end{eqnarray}
(here we use that $\varepsilon_{-1}<0$ and $\varepsilon_{1}>0$). In the limit $\kappa\to 0$ the dependences
on the $\varepsilon$ and $\xi$ factorizes, and we get
\begin{equation} 
   W(\varepsilon_{-1}, \varepsilon_1) = 
   {\rm const}\cdot (\varepsilon_{1}^2|\varepsilon_{-1}|+\varepsilon_{1}|\varepsilon_{-1}|^2)
   g(\varepsilon_{-1}) g(\varepsilon_{1})
\end{equation}

Integrating over $\varepsilon_{-1}$ we get the two-level spacing distribution
\begin{equation} 
  \frac{\mathd w}{\mathd s} = f(s)\equiv C\; (a_1 s^2+a_2 s) g(s), \qquad\qquad 
   a_k=\int_{0}^{\infty} \mathd \varepsilon \; \varepsilon^k g(\varepsilon).
\end{equation}
In correspondence to (\ref{Wsurm}) we choose $g(s)=\exp(-b s)$ where $b$ is some constant to be found 
below. Then $a_2=1/2b$, $a_1=\sqrt{\pi}/(4b^{3/2})$ and the normalization condition reads
\begin{equation} 
  1 = \int_0^\infty \mathd s f(s)=C\sqrt{\pi}/4 b^{5/2},  \qquad\qquad
  C= 4b^{5/2}/\sqrt{\pi}
\end{equation}
The last parameter $b$ is defined by a requirement for the mean spacing to be equal to 1:
\begin{equation} 
  1= \int_0^\infty \mathd s \;s \; f(s)=(1+\pi/4)/\sqrt{\pi b},  \qquad\qquad 
    b = (1+\pi/4)^2/\sqrt{\pi}
\end{equation}
Finally from (\ref{F22}), (\ref{F23}) we get the distribution  of the central particle along the axis $\xi$:
\begin{eqnarray}
P(\xi=\sqrt{x},\kappa)=\frac{\mathd W}{\mathd \xi}= {\rm const}\, \mathcal{A}(x,\kappa)
\exp\left\lbrace-\left( 1+\frac{\pi^2\kappa^2}{4}\right)\frac{x}{2} \right\rbrace 
\,\mathcal{I}_0 (x,\kappa) \mathcal{I}_1(x,\kappa),
\end{eqnarray}
where 
\begin{equation}
\mathcal{A}(x,\kappa)=\left(\frac{\sinh(\pi\kappa/2)}{\pi\kappa/2}\right)^2
\frac{\pi\kappa\sqrt{2x(3+x)}/4}{\sinh(\pi\kappa\sqrt{2x(3+x)}/4)}
\times\frac{\left[1+3x\kappa^2/8\left(1+x^2\kappa^2/4\right)\right]}{\left(1+x^2\kappa^2/4\right)}
\end{equation}
and
\begin{eqnarray}
\mathcal{I}_l (x,\kappa)&=&\int_{-\infty}^{\infty} \mathd \xi \int \mathd \varepsilon\,\varepsilon^l\,
\frac{\varepsilon^2+\frac{\kappa^2}{4}(\xi^2-x)^2}
    {\sqrt{\varepsilon^2+\frac{\kappa^2}{4}(\xi^2+x)^2}}\exp\left\lbrace -\frac{\xi^2}{2}
    \right\rbrace\, g(\varepsilon,\kappa)\nonumber\\
 &&=\int_{-\infty}^{\infty} \mathd \xi \int \mathd \varepsilon\,\varepsilon^l\,
 \frac{\varepsilon^2+\frac{\kappa^2}{4}(\xi^2-x)^2}
     {\sqrt{\varepsilon^2+\frac{\kappa^2}{4}(\xi^2+x)^2}}\exp\left\lbrace -\frac{\xi^2}{2}
     \right\rbrace\, \exp\left\lbrace -\frac{(1+\pi/4)^2}{\sqrt{\pi}}\varepsilon^2\right\rbrace
\end{eqnarray}

\section{Numerical simulations of width distributions.}\label{sec:Num}

Any statistical system with known probability distribution can be easily simulated
numerically, e.g. by Metropolis algorithm \cite{Metro53} (see, for introduction 
the textbook \cite{Binder86}). In particular, applying this methods to joint distribution
 (\ref{eq:distSV})   allows to generate random sets of levels and width that appear 
 with the  probability  corresponding to the given probability distribution (\ref{eq:distSV}).
 
%
 In practical simulations instead of  the original RMT joint distribution (\ref{eq:distSV})
 we use its nonsingular form in terms of scaled variables $\varepsilon_n=E_n/D$ and 
 $\xi_n=\sqrt{x}=\sqrt{\Gamma_n/\langle \Gamma\rangle}$, see eqs(\ref{eq:eps.x}), (\ref{eq:PT-xi}).
 
Of course, in practical simulations the total number $N$ of simulated levels  is 
finite. Moreover, the measured width distribution should belong to levels that are far 
enough from edges in order to avoid the finite-size effects.  
More exactly, we collect the statistics for levels lying in the center of the semi-circle level
distribution, where the mean density of levels is constant within $1\div 3$ per cents.
As the criteria that the boundary effects for selected levels becomes weak we 
consider the independence of width distribution form as a function of the system size $N$.
In practice, for $\kappa\lesssim0.3$ insensitivity to $N$ was found approximately
at $N\gtrsim 20-30$ while at $\kappa \gtrsim 0.5$ width distributions remain sensitive to $N$ even 
at $N\gtrsim (200-300)$.

\end{widetext}
\end{document}